\documentclass{ws-mpla}

\begin{document}

\markboth{T.B. Davies et al.}
{Parametric Instability in Scalar Gravitational Fields}

\catchline{}{}{}{}{}

\title{PARAMETRIC INSTABILITY IN SCALAR GRAVITATIONAL FIELDS}

  \author{\footnotesize TREVOR B. DAVIES$^{1,}$\footnote{t.davies@abdn.ac.uk}                                                                        {,  \footnotesize CHARLES H.-T. WANG$^{1,3,}$\footnote{c.wang@abdn.ac.uk} {,   \footnotesize ROBERT BINGHAM$^{2,3,}$\footnote{bob.bingham@stfc.ac.uk}\\
{\textit{and}  \footnotesize J. TITO MENDON\c{C}A.$^{3,4,}$\footnote{titomend@ist.utl.pt}.}}}}

\address{${^{1}}$ SUPA Department of Physics, University of Aberdeen, Aberdeen, AB24 3UE, UK}
\address{${^{2}}$ SUPA Department of Physics, University of Strathclyde, Glasgow, G4 0NG, UK}
\address{${^{3}}$ STFC Rutherford Appleton Laboratory, Chilton, Didcot, Oxfordshire OX11 0QX, UK}
\address{${^{4}}$ GOLP/ Centro de F\'{\i}sica de Plasmas, Instituto Superior T\'{e}cnico, 1049-001 Lisboa, Portugal}

\maketitle

\pub{Received (Day Month Year)}{Revised (Day Month Year)}

\begin{abstract}
We present a brief review on a new dynamical mechanism for a strong field effect in scalar tensor theory. Starting with a summary of the essential features of the theory and subsequent work by several authors, we analytically investigate the parametric excitation of a scalar gravitational field  in a spherically symmetric radially pulsating neutron star.

\end{abstract}

\ccode{PACS Nos.: include PACS Nos.}

\section{Introduction}	

The first theory of gravity with a scalar gravitational potential $|\hat{\Phi}|$ was that of Newton (1687)$^{1,2}$. This naturally prompted others, including Einstein to attempt to incorporate the scalar field into Special Relativity$^{3,4}$. Although futile, it nevertheless paved the way for the modern  theory of gravitation, General Relativity (GR)$^{5}$. Motivated by searches for a unification of gravity with electromagnetism, Kaluza and Klein (1925)$^{6,7}$ succeeded in providing an aesthetically appealing geometrical theory explaining the origin of the cosmological scalar field due to the presence of extra spatial dimensions in a vacuum, higher dimensional GR.

The idea of a time-dependent gravitational constant $G$, dates back to Mach (1906)$^{8}$ who focussed on the expression $G\sim{R_{v}}/M_{v}$, where $R_{v}$ and $M_{v}$ are the radius and mass of the visible universe respectively. This implied that $G$ varied with the distribution of matter in the universe. It is similar to the expression of the critical density of the universe: $3H^{2}=8\pi{G}\rho_{cr}$  where the Hubble parameter $H$ is defined as $H=\dot{a}/a$ with the scale factor $a(t)$, varying as a power of $t$ if one assumes $R_{v}\sim{t^{-1}}$.  However, the observed density is much smaller than the critical density and the deficit is now called dark matter and dark energy respectively$^{9-12}$.

Dirac (1938)$^{13-16}$ proposed that ratios between fundamental constants should be of the order of unity. He  was convinced that combinations of fundamental constants were related in a natural way if one of the constants was allowed to vary with time. He pursued this idea by taking the ratio of two fundamental forces of nature, electrical and gravity on a standard atomic particle: $\gamma=\frac{e^{2}}{\kappa{m^{2}}}\approx{10^{40}}.$ He later related the ratio of the present mass of the universe to that of the standard atomic mass: $\mu\equiv{M_{u}}/m\approx{10^{80}}$. Finally by using $\mu\approx{t^{2}}$ and $\gamma=t$, he developed a cosmological model in which $\mu$ and $\gamma$ varied with the age of the universe. Later as a preface to inflationary cosmology, these \textit{large number coincidences} led to remarkable results: $\mu/t\gamma\approx10^{0}$ and $\kappa/{M_{u}}\approx{10^{0}}$ respectively.

Jordan(1947)$^{7,17}$ started by taking Kaluza's unified field theory in five-dimensional space with its fifth variable, the constant scalar field, as a function to replace the gravitational constant $\kappa$.  He extracted the scalar field from the original five-dimensional gravito-electromagnetic theory and replaced it with a new four-dimensional interpretation in which the field equations involving a scalar field related to Newton's gravitational constant can be interpreted.

 Brans and Dicke (1960)$^{18,19}$ were impressed by Mach's idea which implied that $\kappa$ was a function of the mass distribution in the universe. By introducing the scalar field $\Phi$ which takes on the role of $\kappa$ as the reciprocal of Newton's gravitational constant and taking some motivation from $1/\kappa=M/R$, it is possible that $1/\kappa$ is a field variable that satisfies a field equation with mass density $\rho$ as its primary source. For a comprehensive review of the history of scalar tensor theories, the reader is referred to refs: [7][13][19][20][21].

Interest in Brans-Dicke theory and scalar tensor theories in general dwindled in the 1970s due to the more and more stringent constraints imposed on them by Solar System experiments. The observational limits can only be satisfied by assuming large values of $|\omega|$ where $\omega$ was a free parameter contained in the theory and therefore expected to be of order unity. The lower bound on $|\omega|$ kept getting larger and larger as the experiments became more and more accurate$^{18}$. By the mid 1980s, scalar tensor theories regained a surge of interest mainly due to the importance of scalar fields in modern unified field theories and inflationary cosmology$^{7,21}$.

In the astrophysical context, Damour and Esposito-Far\`{e}se (1993)$^{22-24}$ reported that neutron star models within scalar tensor theories may undergo a phase transition that consists of the appearance of a non-trivial configuration of the scalar field $\Phi$ in the absence of sources complete with vanishing asymptotic value. This phenomena dubbed \textit{spontaneous scalarization}, arises under certain conditions where the appearance of the scalar field gives rise to a configuration that minimizes the star's energy (its ADM mass) with fixed baryon number. It appears even when the parameters of the theory satisfy the stringent bounds placed by the solar system experiments or uniquely even when the Brans-Dicke parameter of the theory is arbitrarily large. This suggests ultimately that weak field effects cannot constrain the effects of the scalar field in the strong field regime and prompts for alternative measurements.

 Sotani and Kokkotas (2005)$^{25,26}$ showed that the presence of a scalar field in a neutron star affects its equilibrium configuration and consequently its oscillation spectrum. These carry clear imprints of the presence of the scalar field.  Observations of the neutron star's oscillation spectrum via gravitational waves or electromagnetic signals emanating from or around its surface will not only probe the existence of the scalar field but it might also provide a measurement of its asymptotic value.

Finally Wang, Bonifacio, Bingham and Mendon\c{c}a(2009)$^{27,28}$ proposed a new strong-field effect due to the relaxation of a more general function $a(\Phi)$ to its local minimum during the cosmological evolution. It occurs in extreme conditions with strong time-varying gravity such as the interior of a newly-born neutron star. In this case the scalar gravitational field may be simulated according to parametric instability (\S 5) and
therefore provides an initial estimate of the effect that can be extended
for further investigation with realistic stars, including the
possible energy transfer from a collapsed star core to stalled shock
waves in supernova formations and other astrophysical problems
$^{29,30}$. A further motivation  is to seek a possible source
of conformal fluctuations of spacetime as a result of background
scalar gradational waves $^{27}$.

This review is organized as follows: In \S 2. we discuss the dynamical elements of the theory giving the actions and field equations in two conformal frames complete with an overview of the basic components of the analysis. In \S 3. we discuss the post-Newtonian limit along with tests for gravity in the weak field regime before commenting on the conditions necessary for spontaneous scalarization to occur in neutron stars. In \S 4 and \S5. we derive the field equations for a spherically symmetric neutron star and approximate the quasi normal modes of the scalar field to that of a damped harmonic oscillator. In \S 6. we adopt the method used by Wang et al.$^{28}$ to simulate the parametric excitation of scalar fields in a proto-neutron star inducing  strong field effects which we later analyze using  stability methods.

\section{The Equations of Scalar Tensor Theory}

 Jordan-Brans-Dicke theory (BD hereafter) is the prototypical Scalar Tensor Theory of gravity$^{7,21}$. The action in the Jordan frame takes the form \begin{equation}S_{BD}=\frac{1}{16\pi}\int{d^{4}x}\sqrt{-g}\left[\Phi{R}-\frac{\omega(\Phi)}{\Phi}g^{ab}\nabla_{a}\Phi\nabla_{b}\Phi-V(\Phi)\right]+S_{m}\end{equation} where \begin{equation}S_{m}=\int{d^{4}}\sqrt{-g}\mathcal{L}_{m}\left(g_{ab},\Psi\right).\end{equation} The gravitational field is described by the Jordan frame metric $g^{ab}$ and the BD scalar $\Phi$  which along with matter variables $\Psi$ describes the field dynamics. $R$ is the Ricci curvature scalar formed from $g^{ab}$ and the BD parameter $\omega$ is rendered dimensionless by the denominator $\Phi$ in the second term in the action. The Lagrangian density $\mathcal{L}_{m}$ does not depend on $\Phi$ minimally coupling to matter, instead $\Phi$ is directly coupled to the Ricci curvature scalar $R$. The scalar field potential $V(\Phi)$ generalizes the cosmological constant and is often used in inflationary theories of the early universe and present-day quintessence$^{31-33}$.  We work with dimensionless units $(G=c=1)$ and adopt the metric signature $(-+++).$ We rewrite the generalized Einstein equations in the Jordan frame by varying $S_{BD}$ with respect to $g^{ab}$ obtaining \begin{equation}\delta\left(\sqrt{-g}\right)=-\frac{1}{2}\sqrt{-g} g_{ab}\delta g^{ab}\end{equation} and
\begin{equation}\delta(\sqrt{-g}R)=\sqrt{-g}\left(R_{ab}-\frac{1}{2}g_{ab}R\right)\delta g^{ab}\equiv \sqrt{-g} G_{ab}\delta g^{ab}\end{equation} leading to \begin{equation}G_{ab}=\frac{1}{\Phi}\left(\nabla_{a}\nabla_{b}\Phi-g_{ab}g^{cd}\nabla_{c}\nabla_{d}\Phi\right)-\frac{V}{2\Phi}g_{ab}+\frac{\omega}{\Phi^{2}}\left(\nabla_{a}\Phi\nabla_{b}\Phi-\frac{1}{2}g_{ab}\nabla^{c}\Phi\nabla_{c}\Phi\right)+\frac{8\pi}{\Phi}T_{ab}\end{equation} where \begin{equation}T_{ab}\equiv\frac{-2}{\sqrt{-g}}\frac{\delta}{\delta g^{ab}}(\sqrt{-g}{\mathcal{L}})\end{equation} is the energy-momentum tensor for ordinary matter $\Psi$. Varying the action with respect to the scalar field $\Phi$ yields: \begin{equation}\frac{2\omega}{\Phi}g^{cd}\nabla_{c}\nabla_{d}\Phi+R-\frac{\omega}{\Phi^{2}}\nabla^{c}\Phi\nabla_{c}\Phi
-\frac{dV}{d\Phi}=0.\end{equation} By taking the trace of the Einstein equation in the Jordan frame we obtain \begin{equation}R=\frac{-8\pi{T}}{\Phi}+\frac{\omega}{\Phi^{2}}\nabla^{c}\Phi\nabla_{c}\Phi+\frac{3g^{cd}\nabla_{c}\nabla_{d}\Phi}{\Phi}+\frac{2V}{\Phi}\end{equation}
and by eliminating $R$ we derive the scalar field equation in the Jordan frame \begin{equation}\Box\Phi=\frac{1}{2\omega+3}\left[8\pi{T}+\Phi\frac{dV}{d\Phi}-2V\right]\end{equation} where $\Box\Phi=g^{cd}\nabla_{c}\nabla_{d}\Phi$ is the Laplace-Beltrami operator of $g_{cd}$.
  The conservation equation $\nabla_{a}T^{ab}=0$ regulating the dynamics of this matter is conformally invariant and $T\doteqdot{g^{ab}}T_{ab}$ represents the trace of the vanishing energy momentum tensor. Converting from the Jordan frame to the Einstein frame is equivalent to converting from a frame in which the scalar field is non-minimally coupled to the metric tensor over to a frame where the scalar field is minimally coupled to the metric tensor$^{34-36}$. To make the conversion, consider a spacetime $(\mathcal{M},g_{ab})$ were $\mathcal{M}$ is a smooth $D$-dimensional manifold and $g_{ab}$ is a Lorentzian metric on $\mathcal{M}$. The following conformal transformations \begin{equation}\tilde{g}_{ab}=\mathcal{A}^{2}(\Phi)g^{*}_{ab}\end{equation} \begin{equation}\tilde{g}^{ab}=\mathcal{A}^{-2}(\Phi)g^{ab}_{*}\end{equation} \begin{equation}\sqrt{-\tilde{g}}=\mathcal{A}^4(\Phi)\sqrt{-g}_{*}\end{equation} \begin{equation}\tilde{T}_{ab}\equiv\mathcal{A}^{2}(\Phi)T^{*}_{ab}\end{equation} \begin{equation}\tilde{R}=\mathcal{A}^{-2}(\Phi)\left[R_{*}-2(D-1)\frac{\Box{\mathcal{A}}}{\mathcal{A}}-(D-1)(D-4)g^{ab}_{*}\frac{\mathcal{A}_{,a}\mathcal{A}_{,b}}{\mathcal{A}^{2}}\right]\end{equation}  \begin{equation}\tilde{\Box}\Phi=\mathcal{A}^{-2}(\Phi)\left(\Box_{*}\Phi+(D-2)g^{ab}_{*}\frac{\mathcal{A}_{,a}}{\mathcal{A}}\Phi_{,b}\right)\end{equation} are derived where $\mathcal{A}(\Phi)$ is a smooth, non-vanishing function of the spacetime point in a point-dependent rescaling of the metric. It is called a \textit{conformal factor} and must have values which lie within the range $0<\mathcal{A}<\infty$ (\textsl{a, b, k, l=0, 1, 2}....D). All starred quantities represent components in the conformally transformed Einstein frame while quantities with tilde are components in the Jordan frame. These transformations may stretch or shrink distances between points described by the same coordinate system on the manifold but the angles between the vectors is always preserved leading to a conservation of the global causal structure. Computations presented in the
literature are performed in the Einstein frame, because it leads to well posed Cauchy
problems (that is elliptic and/or hyperbolic equations with a set of initial conditions) with perfectly regular dynamics$^{37,38}$. The cosmological evolution resulting from
these computations can be later expressed in the Jordan frame, where the interpretation
of the observable quantities is easier.
Conformal transformations are simply localized scale transformations where $\mathcal{A}=\mathcal{A}(x)$. In conformally flat spacetimes of the form $\tilde{g}_{ab}\mathcal{A}^{2}=\eta_{ab},$ we  obtain the flat Minkowski metric corresponding to a value of $\eta$. For an authoritative account on conformal transformation in theories of gravity, the reader is referred to refs:[39][40][41].

Finally the general action for the scalar gravitational field in the Einstein frame takes the form \begin{equation}S=\frac{c^{4}}{16\pi{G_{*}}}{\int}\frac{d^{4}x}{c}\sqrt{g_{*}}R_{*}+S_{\Phi}+S_{m}\end{equation} where $R_{*}$ is the Ricci curvature scalar in the new frame. The specific action for the scalar field $S_{\Phi}$ in terms of a potential function $V(\Phi)$ together with the specific action for the matter field $S_{m}$ in terms of a coupling function $\mathcal{A}^{2}(\Phi)$ is written as \begin{equation}S=-\frac{c^{4}}{4\pi{G_{*}}}{\int}\frac{d^{4}x}{c}\sqrt{g_{*}}\left[\frac{1}{2}g_{*}^{ab}\Phi_{,a}\Phi_{,b}+V(\Phi)\right]+S_{m}\left[\Psi,\mathcal{A}^{2}(\Phi)g^{*}_{ab}\right].\end{equation} All quantities with asterisks are related to the Einstein two-spin metric $g_{ab}^{*}$. They are the bare gravitational constant $G_{*}$ and  the scalar field $\Phi$ with  its self-interaction term $V(\Phi)$ and its coupling to matter $\mathcal{A}(\Phi)$. The functional $S_{m}\left[\Psi,\mathcal{A}^{2}(\Phi)g^{*}_{ab}\right]$ stands for the action of any field $\Psi_{m}$ that contributes to the energy content of the universe. It expresses the fact that all these fields couple universally to a conformal metric $\tilde{g}_{ab}=\mathcal{A}^{2}(\Phi)g_{ab}^{*}$ implying that the weak equivalence principle (the local universality of free fall for non-gravitationally bound objects) holds in this class of theories. The effective energy momentum tensor for $\Phi$ is derived from (17) and has the form \begin{equation}T^{ab}_{\Phi}\doteqdot{2}c\sqrt{-g_{*}}\frac{\delta{S}_{\Phi}}{\delta{g}^{*}_{ab}}= \frac{c^{4}}{8\pi{G}}\left[2g^{ac}_{*}g^{bd}_{*}\Phi_{,c}\Phi_{,d}-g^{ab}_{*}(g^{cd}_{*}\Phi_{,c}\Phi_{,d}+2V(\Phi))\right].\end{equation} The field equation for the scalar field can be obtained by varying the total action in (16) which leads to \begin{equation}\square_{*}{\Phi}-\frac{\partial{V}(\Phi)}{\partial{\Phi}}=-\frac{4\pi{G}_{*}}{c^{2}}a(\Phi)T_{*}.\end{equation} The cosmological scale factor $a(\Phi)$ and the field derivatives $\alpha(\Phi)$ and $\beta(\Phi)$ are related to the conformal factor $\mathcal{A}(\Phi)$ by \begin{equation}a(\Phi)=ln\mathcal{A}(\Phi)\end{equation} \begin{equation}\alpha(\Phi)=\frac{dln\mathcal{A}(\Phi)}{d\Phi}\end{equation} \begin{equation}\beta(\Phi)=\frac{d^{2}ln\mathcal{A}(\Phi)}{d\Phi^{2}}.\end{equation}In this review we are interested in values where $\Phi$ is near a local \emph{minimum} of $a(\Phi)$. Thus up to an additive constant, equivalent to a re-scaling constant for the metric $g_{ab}$, we have approximately $a(\Phi)=\frac{1}{2}\beta\Phi^{2}$ for some constant $\beta>0$. For simplicity we adopt the quadratic potential $V(\Phi)=\frac{1}{2}\mu^{2}_{0}\Phi^{2}$ which gives the effective mass $m_{0}=\mu_{0}\hbar/c$ of the scalar field $\Phi$ in vacuum$^{28}$. The scalar field equation thus becomes \begin{equation}\square_{*}\Phi-\mu^{2}_{0}\Phi=U\Phi\end{equation} where \begin{equation}U\doteqdot-\frac{4\pi{G}_{*}\beta}{c^{4}}T_{*}\end{equation}

\section{Strong-field effects and spontaneous scalarization}

 A key motive for studying scalar tensor theories of gravity is the strong desire to embed GR into a class of consistent alternatives. However, there is an increasing need to test certain features of the theory for consistency, completeness and to check for agreement with past experiments$^{42}$. Recall that the action of geometry on matter in scalar tensor theories is the same as in GR, but that the dynamics of geometry and the action of matter on it is modified because of the presence of the scalar field.

In the post-Newtonian formalism (PN), the analysis of Solar System tests in the weak field regime for any metric theory of gravity can be simplified using an expansion of the small parameters: $|\hat{\Phi}|$, $\Pi$,  $v^{2}$ and $|T_{jk}|/{\rho_{0}}$ respectively $^{43}$.  $|\hat{\Phi}|$ represents the Newtonian potential and $\Pi$ is the internal density per unit baryon mass density. The parameter $v^{2}$ is the square of the velocity relative to the Solar System centre of mass while $|T_{jk}|/{\rho_{0}}$ is the stress per baryon mass density. The baryon mass density is simply a measure of the number density of baryons $n$. Such corrections give the Newtonian treatment of the Solar System in first order and the  post-Newtonian corrections to the Newtonian treatment in second order$^{44}$.

The parameterized post-Newtonian formalism (PPN) is a calculational tool used for all metric theories of gravity to explicitly express the parameters in which a theory of gravity can differ from Newtonian gravity. One set of values for these parameters makes the PPN formalism identical to the PN limit for GR, while another set of values makes the formalism identical to BD theory etc. Metric theories of gravity only differ from each other in the way their laws generate the metric.

 It is widely acknowledged that GR breaks down at the limit of strong gravitational fields$^{7,43}$. Consequently, when one considers the theory as a classical geometric description of spacetime, it yields predictions of infinite densities and curvatures in the formation of blackholes with a singularity at its centre. This situation persists  even when integrating backwards in time in the evolution of a uniform and isotropic universe. Quantum gravity prohibits such unphysical solutions that occur at the limit of infinitely strong gravitational fields. Recent developments that promise to test the strong-field regime can allow us to place constraints on deviations from GR that are as large as $\sim10$ orders of magnitude more stringent compared to existing tests which have all been done on the Solar System$^{45,46}$.  The strongest gravitational field in the Solar System is that of the Sun which corresponds to a spacetime curvature of \begin{equation}\frac{GM_{\odot}}{R_{\odot}^{3}c^{2}}\simeq{4\times10^{-28} cm^{-2}}\end{equation} and a gravitational redshift of \begin{equation}z_{\odot}\backsimeq\frac{GM_{\odot}}{R_{\odot}c^{2}}\simeq{2\times10^{-6}}.\end{equation}  These are substantially weaker fields compared to that found in the vicinity of  neutron stars and stellar mass blackholes which have a spacetime curvature of $\simeq{2\times10^{-13}}cm^{-2}$ and a gravitational redshift of $\sim1$. In light of this, there is no reason why the equations of GR must be chosen over alternatives. A self-consistent metric theory of gravity can be constructed for any other action as long as it can reproduce the Minkowski spacetime in the absence of matter fields and the cosmological constant. It must also be constructed from only the Riemann curvature tensor and metric and  must follow the symmetries and conservation laws of the energy momentum tensor of matter. Finally, it must be able to produce Poisson's equation in the Newtonian limit$^{43}$.

The strength of a gravitational field at a distance $r$ away from an object of mass $M$ is measured by the parameter \begin{equation}\epsilon\equiv\frac{GM}{rc^{2}}\end{equation} which is proportional to the Newtonian gravitational potential and is directly related to the redshift$^{45}$. Infinitesimal gravitational fields correspond to the limit $\epsilon\rightarrow0$,  leading to the Minkowski spacetime of special relativity. Weak gravitational fields correspond to $\epsilon\ll1$,  leading to Newtonian gravity. Finally, strong gravitational fields are characterised by $\varepsilon\rightarrow1$ at which point the blackhole horizon of an object of mass $M$ is approached.

At higher post-Newtonian orders $1/c^{2}$, any deviation from GR involves at least two factors  and has the form \begin{equation} \mathcal{\textit{\textbf{Z}}}= \alpha^{2}_{0}\times\left[\lambda_{0}+\lambda_{1}\frac{Gm}{Rc^{2}}+\lambda_{2}(\frac{Gm}{Rc^{2}})^{2}+....\right]\end{equation} where $\mathcal{\textit{\textbf{Z}}}$ is the deviation from GR and $\alpha_{0}$ is a constant related to the BD parameter $\omega_{BD}$ by $\alpha^{2}_{0}= (2\omega_{BD}+3)^{-1}.$  The most stringent bound on $\omega_{BD}$ was provided by the Cassini Spacecraft suggesting that $\omega_{BD}> 40000$  implying that the larger $\omega_{BD}$ gets, the weaker the scalar field coupling. $R$ and $m$ denote the radius and mass of the the body under consideration and $\alpha_{0}$, $\alpha_{1}$......are known constants built from the coefficients $\alpha_{0}$, $\beta_{0}$ from the expansion \begin{equation}ln{A}(\Phi)\equiv\alpha_{0}(\Phi-\Phi_{0})+\frac{1}{2}\beta_{0}(\Phi-\Phi_{0})^{2}+\mathcal{{O}}(\Phi-\Phi_{0})^{3}\end{equation} derived at the background value $\Phi_{0}$ of the scalar field: \begin{equation}\gamma^{PPN}-{1}=-\frac{2\alpha^{2}_{0}}{1+\alpha^{2}_{0}}\end{equation}  \begin{equation}\beta^{PPN}-{1}=\frac{1}{2}\frac{\alpha_{0}\beta_{0}\alpha_{0}}{(1+\alpha^{2}_{0})^{2}}.\end{equation}  The Eddington parameters($\gamma$ and $\beta$) are related by $\beta^{PPN}=\gamma^{PPN}=1$ for GR but these parameters can differ for scalar tensor theories$^{46}$. The factor $\alpha^{2}_{0}$ is experimentally known to be small and expected to be close to GR at any order$^{44}$.  It is obtained from the exchange of a scalar particle between two bodies, whereas $\alpha_{0}\beta_{0}\alpha_{0}$ comes from a scalar exchange between three bodies. For a comprehensive review of the PN approximation for relativistic compact binaries, the reader is referred to refs:[47][48][49].

Some non-perturbative effects may occur in strong field conditions if the compactness $Gm/Rc^{2}$ of a body is greater than a critical value $\Phi_{c}.$ This is notable for neutron stars whose compactness are of order $Gm/Rc^{2}\sim0.2$,  compared to $2\times10^{-6}$ for the Sun. There is no deviation from GR at any order in a perturbative expansion in powers of $1/c$. Using a simple parabolic function of the form \begin{equation}A_{\beta}(\Phi)\equiv{exp}\left(\frac{1}{2}\beta_{0}\Phi^{2}\right),\end{equation} the scalar field at the centre of a static body takes a particular value $\Phi_{c}$ which decreases as $1/c$ outside. Harada (1998)$^{50}$ reported that when the condition \begin{equation}\beta_{0}\equiv\frac{\partial^{2}lnA(\Phi_{0})}{\partial\Phi^{2}_{0}}\leq-4\end{equation} is satisfied, the function $\Phi_{c}$ has the shape of a Mexican hat$^{46}$ giving the value $\Phi_{c}=0.$ This represents a local \emph{maximum} where it is energetically favourable for the compact object to create a non-vanishing scalar field and thereby a non-vanishing scalar charge. The coupling strength ${\alpha}(\Phi)=\partial{lnA}(\Phi)/\partial{\Phi}=\beta_{0}\Phi_{c}$ generates non-perturbative strong field effects in the compact object which induces order-of-unity deviations from GR. This phenomena is known as \textit{spontaneous scalarization}$^{51,52}$ in analogy with the spontaneous magnetization arising in ferromagnets below the Curie temperature$^{53}$.

\section{Coordinates and metric for a static, spherically symmetric neutron star}

The metric describing a non-rotating, unperturbed, spherically symmetric neutron star modeled as a self-gravitating fluid of cold degenerate matter at equilibrium takes the form \begin{equation} ds^{2}=-e^{2\psi}dt^{2}+e^{2\Lambda}dr^{2}+r^{2}\left(d\theta^{2}+\sin^{2}\theta{d}\phi^{2}\right)\end{equation} where  $\psi=\psi(r)$ and $\Lambda=\Lambda(r)$. The solution to the BD field equation  below is one of four  solutions and is most frequently used in the literature as it is valid for all values of $\omega_{BD}$. \begin{equation}e^{\psi}=e^{\psi0}\left[\frac{1-\frac{B}{r}}{1+\frac{B}{r}}\right]^{k}\end{equation} \begin{equation}e^{\Lambda}=e^{\Lambda0}\left(1-\frac{B}{r}\right)^{2}\left[\frac{1-\frac{B}{r}}{1+\frac{B}{r}}\right]^{\frac{(k-1)(k+2)}{k}}\end{equation} \begin{equation}\phi=\phi_{0}\left[\frac{1-\frac{B}{r}}{1+\frac{B}{r}}\right]^{\frac{(1-k^{2})}{k}}\end{equation} where $k^{2}=\frac{(4+2\omega)}{(3+2\omega)}$ and $B$, $\psi_{0}$, $\Lambda_{0}$, $\phi_{0}$ are constants. For an authoritative account on the structure of neutron stars in scalar tensor theory see refs:[25][26].

We adopt the equation of state used by Morganstein et al(1967)$^{54}$ for the contracted stress energy tensor  \begin{equation}T_{*} = -c^{2}\rho+{3p}\end{equation} where $\rho$ is the density and $p$ the pressure in a star of radius $R$.  We assume the inequality $c^{2}\rho\gg{p}$ so that (24) reduces to \begin{equation}U=\frac{4\pi{G_{*}}\beta}{c^{2}}\rho.\end{equation} For simplicity, we approximate to a flat spacetime corresponding to the Minkowski metric $\eta_{ab}$  leading to the scalar field equation in the form \begin{equation}\partial^{2}\Phi-\triangle\Phi+\mu^{2}_{0}\Phi+U\Phi=0\end{equation} where $\triangle$ is the 3-dimensional Laplace operator. The density fluctuations of the star become $\partial^{2}_{0}\rho-(v/c)^{2}\triangle\rho=0$ where $v$ is the speed of the pressure/density wave and the surface density for a single mode radial oscillation is $r=R.$  Fluctuations of the form \begin{equation}\rho=\rho_{0}[1-\epsilon\chi_{m}(r)\cos(\Omega_{m}t)]\end{equation} are generated where \begin{equation}\Omega_{m}=\frac{m\pi{v}}{R}\end{equation} is the mode index, $\epsilon$ the amplitude parameter and the function \begin{equation}\chi_{n}(r)\simeq\frac{R}{r}\sin(\kappa_{n}r)\end{equation} contains $\kappa_{n}={n}\pi/R$ which is the wave number when $\textit{n=1, 2,....}$ The orthogonality relation is  \begin{equation}\int^{R}_{0}{dr}r^{2}\chi_{n}(r)\chi_{m}(r)=\frac{R^{3}}{2}\delta_{nm},\end{equation} and when substituted into $U=4\pi{G_{*}}\beta/c^{2}\rho,$ the density fluctuations yield \begin{equation}U=U_{0}[1-\epsilon\chi_{m}(r)\cos(\Omega_{m}t)].\end{equation} The scalar field equation now takes a new form \begin{equation}\partial^{2}\Phi-\triangle\Phi+\mu^{2}_{0}\Phi-\epsilon{U_{0}}\chi_{m}(r)\cos(\Omega_{m}t)\Phi=0\end{equation} where $U_{0}\simeq4\pi{G_{*}}\beta\rho_{0}/c^{2}\geq0$ and $\mu^{2}\simeq\mu^{2}_{0}+U_{0}$. Finally the scalar field potential becomes \begin{equation}V(\Phi)=\frac{1}{2}\mu^{2}\Phi^{2}\end{equation} and \begin{equation}V(\Phi)=\frac{1}{2}\mu^{2}_{0}\Phi^{2}\end{equation}  inside and outside the star respectively.

\section{Normal modes of the scalar field inside the neutron star}

The scalar field inside the star can be approximated to a standing wave subject to the boundary conditions $\Phi(R,t)=0$ due to the surface $r=R$ behaving like an anti-phase reflector for outgoing $\Phi$. Under the conditions $\mu/\mu_{0}\gg{1}$ and $\epsilon=0$, the normal mode of the scalar field becomes \begin{equation}\Phi\approx\sum_{n}\Phi_{n}\doteqdot\sum_{n}\varphi_{n}(t)\chi_{n}(r).\end{equation}  Each $n$ represents a normal mode as \begin{equation}\varphi_{n}t=\Re\varphi_{n0}e^{-i\omega_{n}t}\end{equation} where $\varphi_{n0}$ is the modal amplitude constant giving the energy of $\Phi_{n}$ as \begin{equation}E_{n}=4\pi\int^{R}_{0} {dr} r^{2}u=\frac{c^{2}}{2{G_{*}}}R^{3}\varphi^{2}_{n0}\omega^{2}_{n}\end{equation} where $u=\frac{c^{4}}{8\pi{G_{*}}}[(\Phi,0)^{2}+(\Phi,r)^{2}+\mu^{2}\Phi^{2}]$ is the energy density of $\Phi$ inside the star in spherical coordinates.

 Under certain conditions, the star's surface does allow some scalar wave to propagate through it. When $\mu/\mu_{0}>1,$ there is a loss of energy in which case $\Phi_{n}$ is approximated to quasi-normal modes. We assume that (49) is valid over a few circles of oscillation at angular frequency $\omega_{n}$. Exterior to the star $(r>R)$, this yields  \begin{equation}\varphi_{n}t=\Re\varphi_{n0}\frac{\kappa_{n}}{k_{n}}\frac{R}{r}e^{i(k_{n}r-\omega_{n}t+\theta_{n})}\end{equation} where $\theta_{n}$ is a constant phase and \begin{equation}k^{2}_{n}=\frac{\omega^{2}_{n}}{c^{2}}-\mu^{2}_{0}\end{equation} while \begin{equation}\frac{\kappa^{2}_{n}}{k^{2}_{n}}=\frac{\kappa^{2}_{n}}{\kappa^{2}_{n}+U_{0}.}\end{equation}
 The power carried by the outgoing waves at ${r}\gg{R}$ is \begin{equation}P_{n}={4}\pi r^{2}|{f^{\beta}}|=\frac{c^{4}}{G_{*}}\frac{\kappa^{2}_{n}}{k^{2}_{n}}R^{2}\varphi^{2}_{0}\omega_{n}k_{n}\end{equation} from equation (18) where $f^{\beta}\doteqdot{c}T^{0\beta}_{\Phi}=-\frac{c^{5}}{4\pi G_{*}}\Phi_{,0}\Phi_{,\beta}$ is the flux density of $\Phi.$ At this stage the damping factor \begin{equation}d_{n}=\frac{P_{n}}{E_{n}}=\frac{2c^{2}k^{2}_{n}}{R\omega_{n}k_{n}}\end{equation}  can be obtained for $\varphi_{n}$   and a quasi-normal mode satisfying the damped oscillator equation is described by \begin{equation}\frac{d^{2}\varphi_{n}}{dt^{2}}+d_{n}\frac{d\varphi_{n}}{dt}+\omega^{2}_{n}\varphi_{n}=0.\end{equation}
 For an authoritative account of quasi-normal modes in scalar tensor theory the reader is referred to refs:[25][55][56].

\section{Parametric excitation of normal modes}

Non-linear oscillating systems consist of two or even more subsystems, where one of them is excited; the primary system and the other ones are coupled through non-linear terms and are forming the secondary or excited system. The primary system is an oscillator which can be excited externally, parametrically or by self-excitation, while the secondary system is excited indirectly through the non-linear coupling. In the presence of the density oscillation: $\rho=\rho_{0}[1-\epsilon\chi_{m}(r)\cos(\Omega_{m}t)]$ with $\epsilon\neq0$ and neglecting mode coupling while incorporating the damping factor, parametric instability can be simulated.  We adopt the method used by Wang et al$^{28}$ by applying (49) with a single mode for some $n$ into (46) and then extracting the equation for $\varphi_{n}$  using $\chi_{n}$ as a test function for each quasi-normal mode.  Applying the orthogonality relation \begin{equation}\int^{R}_{0}{dr}r^{2}\chi_{n}(r)\chi_{m}(r)=\frac{R^{3}}{2}\delta_{nm}\end{equation}for any $\textit{m, n = 1, 2,.....}.$ we   obtain \begin{equation}\int^{R}_{0}{dr}r^{2}\chi_{n}[(\partial^{2}_{0}\varphi_{n})\chi_{n}-\varphi_{n}\bigtriangleup\chi_{n}+\varphi_{n}\mu^{2}\chi_{n}-\epsilon{U_{0}}\chi_{m}(r)cos(\Omega_{m}t)\varphi_{n}\chi_{n}]=0\end{equation} which yields \begin{equation}\frac{d^{2}\varphi_{n}}{dt^{2}}+d_{n}\frac{d\varphi_{n}}{dt}+\omega^{2}_{n}\varphi_{n}-\epsilon{c^{2}}{U_{0}}\chi_{mn}\cos(\Omega_{m}t)\varphi_{n}=0\end{equation}
  where \begin{equation}\chi_{nm}\doteqdot\frac{2}{R^{3}}\int^{R}_{0}{dr}r^{2}\chi^{2}_{n}\chi_{m}.\end{equation}These oscillatory modes can be further scrutinized by recasting them into the damped canonical Mathieu equation  \begin{equation}\frac{d^{2}\varphi_{n}}{d\tau^{2}}+2\zeta\frac{d\varphi_{n}}{d\tau}+a\varphi_{n}-2q\cos(2\tau)\varphi_{n}=0\end{equation} where their stability can be analysed according to the relations \begin{equation}\tau=\frac{\Omega_{m}}{2}t\end{equation} \begin{equation} \zeta=\frac{2c^{2}\kappa^{2}_{n}}{R\Omega_{m}\omega_{m}k_{n}}\end{equation} \begin{equation}a=\frac{4c^{2}}{\Omega^{2}_{m}}(\kappa^{2}_{n}+\mu^{2}_{0}+U_{0})\end{equation} \begin{equation}q=\frac{2\epsilon{c^{2}}U_{0}\chi_{mn}}{\Omega^{2}_{m}}.\end{equation}

  Recall that the stability domain near the principal parametric excitation frequency of the scalar wave equation takes the form of (46): $$\partial^{2}\Phi-\triangle\Phi+\mu^{2}_{0}\Phi-\epsilon{U_{0}}\chi_{m}(r)\cos(\Omega_{m}t)\Phi=0$$ by setting $a\approx1$ and $q\approx0$ respectively. However when $a=1$ or $\Omega_{m}=2\omega_{n}$, instability occurs for $\varphi_{n}$  if the condition \begin{equation}\left|\frac{q}{2\zeta}\right|=\left|\frac{\epsilon\chi_{mn}R}{4}\frac{U_{0}\sqrt{\kappa^{2}_{n}+U_{0}}}{\kappa^{2}_{n}}\right|\gtrsim1\end{equation} is satisfied. This comes directly from \begin{equation}U_{0}=\frac{4\pi{G_{*}}\beta}{c^{2}}\rho_{0}\gtrsim0\end{equation} for sufficiently large $\beta$ and $\epsilon$.

  As an example, consider a neutron star with amplitude parameter $\epsilon=1/3$, equilibrium density $\rho_{0}=10^{15}gcm^{-3}$ which is radially pulsating with density wave speed $v=0.75c,$ mode index $m=12,$ frequency $\Omega_{m}/2\pi=90kHz$ and radius $R=15km.$ (41)(42). Using (53), this frequency is twice the frequency of the lowest quasinormal mode of a massless scalar field with $\mu_{0}=0$, $n=1$ and $\omega_{n}/2\pi=45kHz.$ From (61) it then follows that $\chi_{nm}=7.5\times10^{-4},$ therefore the estimated  unstable $\beta$ values using (67) yields $\beta\gtrsim1400.$

  \section{Acknowledgements}
We are grateful to A. Murphy (Edinburgh) for helpful discussions and to the STFC  Centre for Fundamental Physics for support.

\end{document}